\documentclass[reprint,amsmath,amssymb,aps,twocolumn]{revtex4}
\usepackage{graphicx}
\usepackage{dcolumn}
\usepackage{bm}
\usepackage{braket}
\usepackage{hyperref}
\usepackage[utf8]{inputenc}

\begin{document}
\title{Cooper Pairing Mediated by Onsite Exchange: A Possible Mechanism For High-Temperature Superconductivity}	
	
\author{Jacques R. Eone II}
\affiliation{%
	Department of physics, University of Yaound\'e  I \\
	P.O. Box 812, Yaound\'e, Cameroon \\
}	
	
\date{\today}
	
\begin{abstract}
Among the various mechanisms proposed for unconventional superconductivity, this paper focuses on the Coulomb interaction responsible for $d$-wave and $s_\pm$-wave pairing symmetries in cuprates and iron pnictides. Although the effective multi-orbital interaction $U_\text{eff}=U-J$ is predominantly repulsive, an attractive component driven by the Hund's coupling parameter $J$ is sufficient to bind fractional charges. This attractive multi-orbital contribution and a repulsive term of the single-band Hubbard model yield the superconducting pairing gap $\Delta_0$ and an estimate of the transition temperature $T_c$. Given the complex electronic structure and vast compositional space of these materials, the model focuses exclusively on the doped superconducting plane hosting these fractional charges. Through this approach, an analytical expression mainly dependent on the Hubbard $U$ and Hund $J$ parameters that reproduces the empirical superconducting dome is derived. Furthermore, the model addresses the characteristic electron and hole doping asymmetry observed in cuprates by accounting for Hund's coupling parameters. Finally, while the model describes strange metal behavior, it currently provides only a qualitative explanation for the pseudogap phase and the underdoped isotope effect.            
\end{abstract}	
	
\maketitle

\section{Introduction}
In conventional Bardeen-Cooper-Schrieffer (BCS) theory, the transition temperature increases monotonically with the density of states at the Fermi surface \cite{bardeen1957}. However, this theory fails to account for unconventional superconductors, where the transition temperature decreases beyond an optimal doping level, resulting in a superconducting dome. Since the discovery of high-temperature superconductivity in cuprates by Bednorz and Müller in 1986 \cite{bednorz1986}, no consensus has been reached regarding the underlying pairing mechanism. Numerous theoretical models have attempted to resolve this fundamental problem, some of them invoking exotic quasiparticles such as excitons and bipolarons \cite{anderson1987,alexandrov1989,Allender1973,berk1966,moriya2000,scalapino2012,raghu2010}. Elucidating the mechanism behind unconventional superconductivity is crucial both for designing optimized materials and for assessing the feasibility of room-temperature superconductivity at ambient pressure. However, modeling these systems from first-principles remains deeply challenging due to the diversity of unconventional superconductors spanning from heavy fermions \cite{steglich1979}, cuprates \cite{bednorz1986}, iron pnictides \cite{kamihara2006,takahashi2008} to recently nickelates \cite{li2019}, as well as the presence of enigmatic pseudogap and strange metal phases \cite{loram1994,bednorz1986}. Furthermore, the fractional stoichiometry of doped cuprates and pnictides, combined with their strongly correlated electronic structures, presents severe theoretical hindrances. While Bednorz and Müller originally invoked a strong electron-lattice interaction as a guiding concept for high-temperature superconductivity, most contemporary theories suggest that phonons may not be the primary mediator of Cooper pairing. In contrast to the retarded, frequency-dependent electron-phonon coupling characteristic of Eliashberg theory \cite{marsiglio2020}, unconventional superconductors are thought to achieve high transition temperatures via instantaneous interactions. The short timescale associated with the larger order parameter results in a short coherence length. Because the superconducting ground state emerges in close proximity to an antiferromagnetic parent compound in both cuprates and iron pnictides, magnetic pairing mechanisms have gained significant traction, with spin fluctuations remaining one of the most prominent candidates for unconventional pairing. Due to the strongly correlated nature of these compounds, most theoretical descriptions of unconventional superconductivity rely on the Hubbard model. In the strong-coupling regime where the hopping parameter $t$ is small compared to the effective Coulomb interaction ($U \gg t$), the $t-J$ model can be derived from the Hubbard Hamiltonian \cite{chao1978}. This model provides an elegant, single-band description that circumvents the explicit separation of the transition metal $d$ and ligand $p$ orbitals. While the three-band model proposed by Emery explicitly accounts for this orbital separation, it does so at the expense of an additional complexity \cite{emery1987}. For cuprates, a low-energy single-band model can be justified from the multi-band model via the formation of Zhang-Rice singlets \cite{zhang1988}. A key advantage of the $t-J$ model is its capacity to naturally capture the pairing symmetry anisotropy, such as $d-$wave pairing in cuprates. However, explaining the asymmetry between $n$- and $p$-type doping in cuprates requires an extended model that includes next-nearest-neighbor interaction terms ($t'$) \cite{leung2003}. While a purely repulsive interaction can serve as the driving mechanism for unconventional superconductivity, a local or effective attractive contribution is ultimately required to consistently bind Cooper pairs. Beyond electron-phonon coupling, very few pairing candidates provide an attractive contribution that is compatible with the constraint of a repulsive interaction capable of inducing gap asymmetry. In this paper, the Hund's coupling-driven pairing is proposed as a viable mechanism for unconventional superconductivity. The single-band model with fractional charges presented in this paper distinctively departs from standard Hund's pairing models found in the literature \cite{roig2022}. Because the repulsive energy contribution scales quadratically with the charge, the pairing mechanism does not require a large Hund's parameter to effectively bind fractional electrons. To validate this single-band approximation, the electronic structure of a single-layer cuprate is explicitly analyzed. Unlike the traditional $t-J$ model, this approach does not explore charge orders such as stripes  \cite{kivelson2003}; instead, it demonstrates that hole doping primarily modifies states at the top of the valence band, whereas electron doping alters the conduction band. By neglecting fluctuations, the repulsive contribution is simplified, yielding a closed-form analytical expression for the superconducting dome that depends mainly on the onsite Coulomb parameter $U$ and the Hund's coupling $J$. Finally, the resulting superconducting order parameter $\Delta_0$ and the corresponding transition temperature $T_c$ are easily derived, demonstrating close alignment with experimental observations.

\section{On the electronic structure of cuprates}
Due to the vast compositional space of cuprate superconductors, the electronic structure analysis is restricted to the isolated CuO$_2$ plane, where the superconducting state originates. Structurally, these materials consist of alternating CuO$_2$ layers arranged in a perovskite-like geometry, separated by metal-oxide spacer layers that act as charge reservoirs. Although we neglect interlayer coupling in this work, it is known to exert an enhancement effect on the transition temperature \cite{tripathi1997,yayeh2023}. Nevertheless, this interlayer coupling is considered weak compared to the dominant intralayer contributions to $T_c$. A qualitative discussion regarding the dependence of the transition temperature on interlayer interactions is provided later in this paper.

\begin{figure}[h!]
	\includegraphics[width=0.95\linewidth]{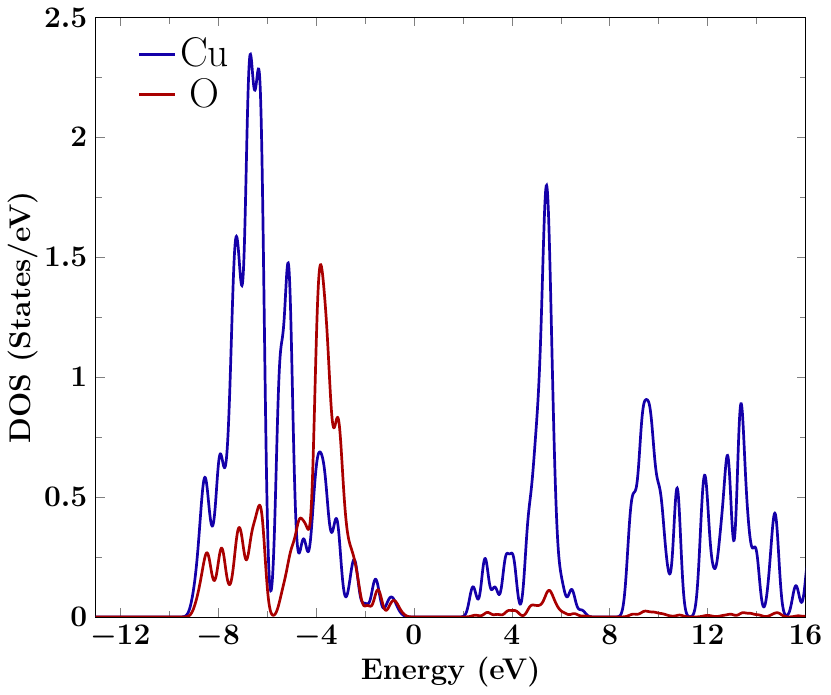}
	\caption{Density of states (DOS) of Cu and O within the CuO$_2$ plane.}
	\label{fig:dos}
\end{figure}

The electronic structure of an undoped CuO$_2$ monolayer can be accurately modeled using first-principles calculations. Fig.~\ref{fig:dos} displays the calculated nonmagnetic density of states (DOS) for Cu and O, obtained via a localized basis-set approach within the local density approximation (LDA) plus Hubbard corrections ($U_d$=5.0 eV for Cu and $U_p$=6.0 eV for O). The resulting band gap is approximately 2 eV, which is comparable to the 2$-$3 eV gaps experimentally reported for CuO$_2$ films \cite{zhong2016,yin2017}. Reflecting a charge-transfer insulator behavior, the top of the valence band is dominated by O states, whereas the bottom of the conduction band is predominantly composed of Cu states. Specifically, these frontier states are governed by the in-plane O$(p_x,p_y)$ and Cu($d_{x^2-y^2}$) orbitals interacting on the square lattice. These bands are not orbitally pure; instead, strong hybridization mixes these states, yielding a valence band with majority-$p$ character and a conduction band with majority-$d$ character. A rigorous description of this hybrid valence state is well-captured by the Zhang-Rice singlet \cite{zhang1988}. Interestingly, this hybridization appears less pronounced in parent compounds such as La$_2$CuO$_4$, where $p$ states heavily dominate below the Fermi level \cite{rivero2010}. This multi-band system can be mapped onto an effective single-band model under the condition that the doped fractional holes (less than 0.5 $h^+$) and fractional electrons (less than 0.5 $e^-$) primarily occupy the O($p$) and Cu($d$) orbitals, respectively. Although this single-band mapping is restricted to the small fractional charge regime, it allows us to effectively describe the dynamics of carriers in both electron-doped and hole-doped cuprates using two distinct single-band Hubbard Hamiltonians:
\begin{align}
H_h &=  -t \sum_{\braket{i,j},\sigma} (p_{i\sigma}^\dagger p_{j\sigma} + \text{h.c.}) + U_p \sum_i n_{i\uparrow}n_{i\downarrow} \nonumber\\
H_e &=  -t \sum_{\braket{i,j},\sigma} (d_{i\sigma}^\dagger d_{j\sigma} + \text{h.c.}) + U_d \sum_i n_{i\uparrow}n_{i\downarrow}, \label{eq:sbhub}
\end{align} 
where $t$ is the hopping parameter, $p_{i\sigma}^\dagger$ ($d_{i\sigma}^\dagger$) creates a hole (electron) at site $i$, and $U_p$ and $U_d$ denote the effective onsite Coulomb repulsions for the $p$ and $d$ orbitals, respectively. The distinct energy scales associated with these carriers are detailed later in this work as a viable solution to the electron-hole doping asymmetry in cuprates. Iron pnictides do not share this specific electronic structure; they are poor metals whose Fermi level is heavily dominated by the Fe($d$) states, while the As($p$) orbitals lie deeper below the Fermi level and hybridize with the Fe($d$) bands \cite{singh2008,Roekeghem2016}. Although iron pnictides are inherently multi-band systems, we effectively reduce the system here to a single-band model governed by a single Coulomb parameter $U_d$ associated with the iron $d$ orbitals.

\section{Pairing in a $d-$wave and $s_\pm-$wave}
\subsection{Symmetry of the order parameter within a tight-binding picture}
\begin{figure}[h!]
	\includegraphics[width=0.95\linewidth]{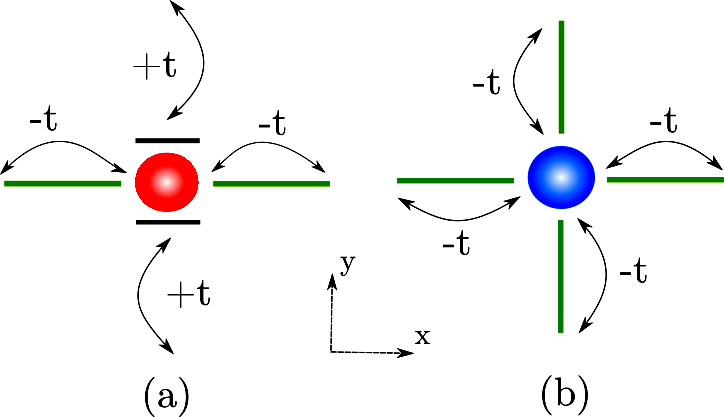}
	\caption{Symmetry of the order parameter within the tight-binding approximation: (a) $d$-wave symmetry in  cuprates and (b) $s_\pm-$wave symmetry in superconducting iron pnictides.}
	\label{fig:x-wave}
\end{figure}

The hopping parameter $t$ in the single-band Hubbard Hamiltonian of Eq.~(\ref{eq:sbhub}) is defined here as positive, without loss of generality due to gauge freedom. Both $d-$wave and $s_\pm-$wave pairing symmetries can be derived within a tight-binding approximation by considering only nearest-neighbor hopping on a square lattice. In this approximation, the tight-binding dispersion relation takes the form:
$$
	\epsilon (k) = -2( t_x\cos k_x + t_y\cos k_y).  
$$
$-\epsilon (k)$ contributes directly to the binding energy. By imposing the condition $t_x=-t_y$, this expression naturally mirrors the momentum dependence of a $d-$wave order parameter:
$$
\Delta(k) = \Delta_0 (\cos k_x - \cos k_y) = -\epsilon (k)|_{t_x=-t_y}, 
$$ 
where $\Delta_0 = 2 t_x = -2 t_y$. This expression physically reflects how the effective hopping matrix elements change sign depending on the orbital orientation and carrier dynamics. This anisotropy is exemplified by the oxygen atoms in the CuO$_2$ plane. While standard longitudinal hopping along the linear Cu$-$O$-$Cu bond (directed along either the $x$ or $y$ axis) is highly favorable, any transverse hopping perpendicular to this bond axis requires overcoming an additional energy barrier, as illustrated schematically in Fig.~\ref{fig:x-wave}a.
Conversely, when hopping is symmetric along both the $x$ and $y$ directions ($t_x=t_y$), as illustrated in Fig.~\ref{fig:x-wave}b, the resulting order parameter exhibits the $s_\pm-$wave symmetry, as observed in the iron pnictides: 
$$
\Delta(k) = \Delta_0 (\cos k_x + \cos k_y) = -\epsilon (k)|_{t_x=t_y}, 
$$ 
with $\Delta_0 = 2 t_x = 2 t_y$. In real space the order parameter takes the form 
\begin{equation}
\Delta_\pm (x,y) = \frac{\Delta_0}{2} [(\delta_{x,-1} + \delta_{x,1})\delta_{y,0} \pm \delta_{x,0}(\delta_{y,-1}+ \delta_{y,1})],
\label{eq:pg_rs}
\end{equation}
where $\Delta_-$ and $\Delta_+$ represent the $d-$wave and $s_\pm-$wave pairing states, respectively. Although this tight-binding picture successfully captures the pairing symmetry of hole-doped cuprates and iron pnictides, it lacks universal generalizability. For instance, if the low-energy electronic structure of electron-doped cuprates were strictly dictated by local interactions at the copper atomic site, this tight-binding approximation would incorrectly predict an $s_\pm-$wave pairing symmetry, while a $d-$wave symmetry is suggested \cite{Chesca2004,Tsuei2000}. Furthermore, several distinct materials with crystal structures entirely unrelated to the cuprates are also predicted to exhibit $d-$wave pairing symmetry \cite{yayeh2023,Izawa2001}. However, this tight-binding model implies that the pairing gap may be driven by a local, onsite attractive parameter that mediates the hopping processes together with competing repulsive contributions giving the resulting order parameter $\Delta_0$.

\subsection{Symmetry of the order parameter due to repulsion}
The pairing order parameter in Eq.~(\ref{eq:pg_rs}) vanishes at the origin (0,0) for both $d-$wave and $s_\pm-$wave symmetries. This spatial node physically manifests the strong onsite repulsive interaction $U$, which prevents Cooper pairing on a single atomic site. In the strong-coupling regime where $U\gg t$, this local constraint allows one to map the system onto the $t-J$ model:
$$
H = - \sum_{\braket{i,j},\sigma} (c_{i\sigma}^\dagger c_{j\sigma} + \text{h.c.}) + J_\text{SE} \sum_{\braket{ij}} \left(\bm{S}_i\cdot\bm{S}_j - \frac{1}{4}n_i n_j\right),
$$ 
where the superexchange parameter is defined as $J_\text{SE}=4t^2/U$. Within this model, pairing is driven by the interaction of electrons occupying nearest-neighbor sites, specifically at ($\pm 1$, $0$) and ($0$, $\pm 1$), which generates a finite pairing gap. The gap equation within BCS theory is given by:
$$
\Delta_k = -\sum_{k'} V_{k,k'} \frac{\Delta_{k'}}{2E_{k'}} \tanh\left(\frac{\beta E_{k'}}{2}\right),
$$
where  $\beta=1/k_\text{B}T$, $k_\text{B}$ is the Boltzmann constant, and $E_k = \sqrt{(\epsilon_k-\mu)^2 +|\Delta_k|^2}$ represents the quasiparticle excitation energy. To satisfy the self-consistent gap equation when the interaction potential is purely repulsive ($V_{kk'} > 0$), the order parameter $\Delta_k$ must undergo a sign change in momentum space, naturally generating an anisotropic pairing symmetry. Consequently, the $t-J$ model serves as a highly robust method for describing superconducting cuprates. However, a notable limitation is its explicit dependence on the superexchange parameter $J_\text{SE}$. In hole-doped cuprates, $J_\text{SE}$ suppresses monotonically with increasing doping, whereas the experimental superconducting transition temperature scales upward with doping until it reaches the optimal doping threshold.

\section{Hund's pairing as a mechanism of unconventional superconductivity}
\subsection{Emergence of Cooper pairing within the single-band Hubbard model}

In a single-band Hubbard model describing $d$- or $p$-orbital states,
$$
H = -t \sum_{\braket{i,j},\sigma} (c_{i\sigma}^\dagger c_{j\sigma} + \text{h.c.}) + U_{d,p} \sum_i n_{i\uparrow}n_{i\downarrow},
$$ 
the repulsive contribution can be treated within a mean-field approximation, where fluctuations are neglected:
$$
(\braket{n_{i\uparrow}}-n_{i\uparrow})(\braket{n_{i\downarrow}}-n_{i\downarrow}) = 0.
$$

This mean-field approximation is justified by the assumption that only a fractional number of electrons $n$ participate in superconductivity and that their mutual interactions $\sim n^2 U$ are therefore weak. Consequently, spin fluctuations are not consistently accounted for within this model. By considering a total number of lattice sites $n_0$, the local magnetic moment is defined as $m = n_0\braket{n_{\uparrow} - n_{\downarrow}}$ and the corresponding number of electrons is given by $n_{d,p} = n_0\braket{n_{\uparrow} + n_{\downarrow}}$ \cite{takahashi2013},
\begin{align*}
U_{d,p} \sum_i n_{i\uparrow}n_{i\downarrow} & \approx U_{d,p} \sum_{k\sigma} n_{k\sigma} \braket{n_{-\sigma}} - n_0 U_{d,p} \braket{n_{\uparrow}} \braket{n_{\downarrow}}\\
&= \frac{1}{2n_0}U_{d,p} \sum_{k\sigma} (n_{d,p} - m \sigma) c_{k\sigma}^\dagger c_{k\sigma}  \\
& \hspace{0.5cm}- \frac{1}{4n_0} U_{d,p} (n_{d,p}^2 - m^2).
\end{align*}
Assuming a non-magnetic superconducting ground state where the local magnetization vanishes ($m=0$), the resulting Hamiltonian centered in an atomic site ($n_0=1$) simplifies to:
$$
H = \sum_{k\sigma}\epsilon_{k} c_{k\sigma}^\dagger c_{k\sigma}  +  \frac{1}{4} U_{d,p} n_{d,p}^2,
$$ 
where the first term represents a single-particle contribution to the band energy, while the second term acts as a two-body repulsive correction. Given that the multi-orbital interaction $U^\text{eff}= U -  J$ is repulsive, the Hund's exchange coupling introduces in this single-band approach as an effective attractive contribution of $-J_{d,p}/4$ within the repulsive term. If this attractive term is treated as a renormalization of the single-particle energy $\epsilon $, thereby modulating the effective hopping parameter $t$, the interaction energy arising from the repulsive component can become effectively attractive for fractional occupations. Driven by the Hund's parameter, this net attraction stabilizes the pairing state and directly determines the magnitude of the superconducting order parameter:
\begin{equation}
E_b =  -\frac{1}{4} J_{d,p} n_{d,p}  +  \frac{1}{4} U_{d,p} n_{d,p}^2 = - \Delta_0.
\label{eq:dome}
\end{equation}
Consequently, it becomes clear that the superconducting state cannot remain stable at integer filling ($n_{d,p}=1$), because the strong onsite Coulomb repulsion dominates the exchange coupling ($U \gg J$). Eq.~(\ref{eq:dome}) provides the analytical expression for the superconducting dome as a function of the carrier concentration ($n_{d,p} \equiv p \equiv n$). Due to the Hund's exchange coupling, the order parameter initially  increases with doping until it reaches an optimal doping threshold, beyond which it is suppressed due to the onsite repulsion. A realistic superconducting dome in unconventional superconductors such as cuprates or iron pnictides never originates at zero doping because of the competing parent antiferromagnetic phase. To capture this physics empirically, the dome curve can be rigidly shifted using an additional offset parameter $p_\text{min}$, yielding the following expression for hole-doped cuprates:

\begin{equation*}
\Delta_0 =  \frac{1}{4} J_{d,p} (p-p_\text{min})  -  \frac{1}{4} U_{d,p} (p-p_\text{min})^2.
\label{eq:dome_h}
\end{equation*}

The corresponding optimal doping level $p_\text{opt}$ and upper critical doping limit $p_\text{max}$ are given by:
$$
p_\text{opt} = \frac{J_{d,p}}{2U_{d,p}} + p_\text{min} \text{  and  }  p_\text{max} = \frac{J_{d,p}}{U_{d,p}} + p_\text{min}.
$$
The maximum order parameter 
$$
\Delta_\text{max} = \frac{1}{16} \frac{J_{d,p}^2}{U_{d,p}},
$$
is related to the maximum transition temperature $T_c$ via the ratio $R = 2\Delta_0 / (k_\text{B} T_c)$. For a $d-$wave pairing symmetry in the weak-coupling limit, this ratio is approximately:
$$
R = \frac{4\pi }{e^\gamma  \sqrt{e} } \approx 4.28,
$$
where $\gamma \approx 0.5772$ represents the Euler constant. Unlike conventional BCS systems, unconventional superconductors lack a single, universally well-defined value for $R$. For instance, values ranging between 4.67 and 6.3 have been suggested for cuprates \cite{Hufner2010}, though even higher ratios are experimentally attainable \cite{Inosov2011}. Furthermore, iron pnictides are multiband superconductors, meaning that the effective ratio $R$ varies significantly depending on whether it is evaluated for the small or large energy gap. In these systems, values of approximately 3.44 or 3.68 are typically reported for a small gap \cite{Gonnelli2009, Chen2009}. However, the weak coupling regime used in this work is satisfied by the condition that $J_{d,p} n_{d,p}$ is weak for $n_{d,p} \ll 1$. Therefore, the values of the gap-to-Tc ratio $R$ are taken as 4.28 and 3.53 for $d-$wave and $s_\pm-$wave symmetries, respectively.

\subsection{Transition temperatures and superconducting dome}
The superconducting order parameter in unconventional superconductors within this theory depends primarily on the Hund's exchange coupling $J$ and the onsite Coulomb interaction $U$. Because these electronic parameters vary across different materials, the order parameter does not exhibit a unique, universal value. Nevertheless, representative parameters can be estimated to characterize both hole- and electron-doped cuprates, as well as iron pnictides. In modern electronic structure theory, the $U$ and $J$ parameters are typically computed from first principles using linear response theory or the constrained random-phase approximation (cRPA) \cite{Cococcioni2005,sasioglu2011,Linscott2018,Moore2024,mcmahan1988}. Alternatively, approximate $U$ values for $p$ and $d$ valence states can be extracted from atomic local density approximation calculations using constrained orbital occupancies \cite{eone2025hubbard}:

\begin{equation*}
	U_{d,p} = \frac{\partial \epsilon_{d,p}}{\partial n_{d,p}},
	\label{eq:u_at}
\end{equation*}

where $\epsilon_{d,p}$ represents the orbital energy of the $d$ or $p$ states. Similarly, the Hund's exchange parameter $J_d$ can be expressed in terms of the Slater integrals $F^2$ and $F^4$, while $J_p$ is derived from the integral $F^2$, defined as: 
$$
F^k = \int_0^\infty  \int_0^\infty r_1^2 r_2^2 \chi_{d,p}^2(r_1)\chi_{d,p}^2(r_2) \frac{r_<^k}{r_>^{k+1}} dr_1 dr_2, 
$$
where $\chi_{d,p}$ represents the localized $d$ or $p$ radial wave function. Representative values of the Hubbard $U$ and Hund's exchange $J$, compiled from literature are employed to evaluate the transition temperature and pairing gap for ideal hole- and electron-doped cuprates, as well as iron pnictides. The resulting parameters are summarized in Table~\ref{tab:table1}. 

\begin{table}[h!]
	\caption{Representative values of the Hubbard $U$, Hund's exchange $J$, and ratio $R$, alongside calculated optimal doping level $p_\text{opt}-p_\text{min}$, zero-temperature maximum pairing gap $\Delta_\text{max}$ and critical transition temperature $T_c$ for a hole-doped cuprate (oxygen state), an electron-doped cuprate (copper state), and an iron pnictide (iron state).}
	\begin{ruledtabular}
		\begin{tabular}{lccc}
			&$p-$doped cuprate & $n-$doped cuprate & Iron pnictide \\
			&O      & Cu   & Fe   \\\hline
		$U$ (eV)& 6.30 \footnote{From Ref \cite{Khn2025}.} & 5.50  \footnote{From Ref \cite{sasioglu2011}.}  & 4.11 \footnote{From  Ref \cite{Moore2024}.} \\ 
		$J$ (eV) & 1.62 \footnote{Upper bound from  Ref \cite{Moore2024}.}& 0.85  \footnote{From Ref \cite{sasioglu2011}.} & 0.80 \footnote{From Ref \cite{Fink2017}.}\\ 	
		$p_\text{opt}-p_\text{min}$ & 0.13 & 0.08 & 0.10 \\ 	
		$\Delta_\text{max}$ (meV)  & 26 & 8 & 10 \\ 	
		$T_c$ (K)& 141  & 44 & 64 \\ 		
		$R $ & 4.28 & 4.28 & 3.53 \\ 	
		\end{tabular}
	\end{ruledtabular}
	\label{tab:table1}
\end{table}

In the weak-coupling limit, the calculated transition temperatures align remarkably well with the highest values experimentally recorded for hole-doped cuprates (about 133 K)\cite{Schilling1993}, electron-doped cuprates (about 40 K) \cite{Er1991,Smith1991}, and iron pnictides (about 55 K)\cite{Ren2008}. Given the intricate electronic structures and large compositional space of these unconventional superconductors, this study focuses on representative configurations optimized for ideal synthesis conditions. Within this single-band approximation, the physical difference between hole- and electron-doped cuprates traces back to the distinct Hund's parameters of the oxygen $p$ and copper $d$ valence states. The experimental pairing gaps often exceed the weak-coupling theoretical values derived in this work, indicating the presence of strong-coupling effects. Furthermore, the ratio $J/U$ lacks a universal value; standard literature estimates generally constrain it within the 0.15 $\le J/U \le$ 0.25 range \cite{Goyal2022,deMedici2011,Georges2013,Zhang2023}. The delicate competition among local interaction parameters ($U$, $J$), chemical doping precision, and specific band structure topologies accounts for the wide diversity of transition temperatures observed in unconventional superconductors. Finally, because the threshold doping concentration required to initiate superconductivity ($p_\text{min}$) is highly material-dependent, it has been omitted from Table~\ref{tab:table1}.

\begin{figure}[h!]
	\includegraphics[width=0.95\linewidth]{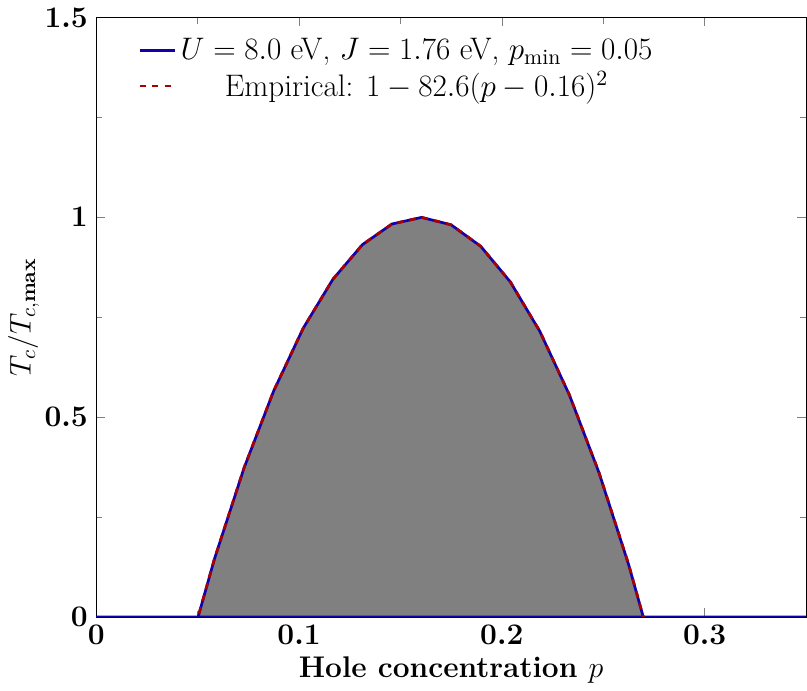}
	\caption{Superconducting dome for a hole-doped cuprate calculated using optimized $U$, $J$, and $p_\text{min}$ parameters, compared to the classic empirical relation of Presland and coworkers \cite{Presland1991}.}
	\label{fig:dome}
\end{figure}

The superconducting dome calculated using optimized values of $U=$8.0 eV, $J=$1.76 eV, and $p_\text{min}=$0.05 is shown in Fig.~\ref{fig:dome}. These parameters are chosen to achieve quantitative agreement with the empirical relation of Presland and coworkers \cite{Presland1991}:
$$
T_c/T_{c,\text{max}}= 1-82.6 (p-0.16)^2.
$$ 
This optimal value of $U=8.0$ eV closely aligns with the value reported for O($p$) in recent literature \cite{Moore2024}, while the corresponding Hund's coupling ($J=1.76$ eV) remains close to the representative parameter listed in Table~\ref{tab:table1}. Within this single-band approximation, these parameters strongly suggest that superconductivity in hole-doped cuprates is fundamentally driven by the electronic structure of the oxygen states rather than that of copper. These parameters yield a dimensionless ratio of $J/U \approx$ 0.22. Evaluating the system in the weak-coupling limit ($R=4.28$) produces a maximum transition temperature of 131 K.\\

Because the superconducting mechanism presented here relies on localized atomic parameters, it might superficially appear to be material-independent. In reality, the Hund's exchange coupling $J$ is sensitive to the surrounding electronic environment, which ultimately dictates the pressure dependence of the transition temperature. High pressure induces a local electronic charge redistribution \cite{Eone2025atomic}, which effectively increases the Hund's parameter as the lattice contracts \cite{Panda2017}; this occurs because a higher local charge density inherently enhances exchange interactions. This pressure-induced enhancement of $J$ with local occupancy offers a qualitative explanation for the observed increase in $T_c$ with the number of coupled CuO$_2$ planes in multilayer cuprates. Conversely, less consensus exists regarding the pressure dependence of the Hubbard $U$. If $U$ undergoes a pressure-induced reduction due to enhanced screening mechanisms \cite{Burnett2024}, the combined effect of an increased $J$ and a decreased $U$ significantly boosts the transition temperature under pressure. Furthermore, the pronounced charge-transfer character of cuprates provides a strong justification for adopting a single-band description in hole-doped configurations. Recently, infinite-layer nickelates have emerged as high-temperature superconductors \cite{li2019,Zeng2020,Osada2020}, albeit with lower transition temperatures than their cuprate counterparts. Although nickelates and cuprates share similar electronic structures, the charge distribution in nickelates, often modeled via a mixed Mott-Hubbard and charge-transfer insulator \cite{bengone2000}, tends to reduce the pairing state. Consequently, the exceptionally high $T_c$ in cuprates can be captured within the theory because their valence states are highly optimized for a single-band projection owing to their robust charge-transfer nature.

\section{About the strange metal behavior, the pseudogap and the underdoped isotope effect}

Assuming a strong repulsive interaction as the primary driving force for superconductivity fails to account for the anomalous properties of the strange metal behavior, because the electron-electron interaction significantly enhances the scattering rate. The strange metal behavior,  characterized by $T$-linear resistivity, is frequently studied in the Planckian limit using the Drude formula \cite{Phillips2022,Patel2023}.   The Drude resistivity
$$
\rho = \frac{m_* }{n e^2 \tau},
$$ 
for a carrier density $n$ of electrons with elementary charge $e$ and effective mass $m^*$, depends inherently on the relaxation time $\tau$. This relaxation time can be expressed as a timescale governed by an energy window $\Delta E$, which is opened by the scattering of a fraction of electrons $n_\text{s} \propto k_\text{B}T/E_\text{F}$, where $E_F$ denotes the Fermi energy. Within the pairing mechanism proposed here, the total scattering energy window features two competing components: A contribution driven by the Hund's coupling, $\Delta E_1  \propto  J n_\text{s}$, and a contribution consistent with Fermi-liquid theory, $\Delta E_2 \propto  U  n_\text{s}^2$. Consequently, the total scattering rate $\tau^{-1}$ is expressed as:
$$
\frac{1}{\tau } \approx \frac{\Delta E}{ \hbar} \propto \left\{
\begin{array}{c}
 \frac{U}{\hbar}    n_\text{s}^2  =  \frac{U}{\hbar}  \frac{ k_\text{B}^2}{E_\text{F}^2} T^2 \text{  for } p \gg p_\text{opt}   \text{ (Fermi liquid)} \\
\frac{J}{\hbar} n_\text{s}   = \frac{J}{\hbar} \frac{ k_\text{B}}{E_\text{F}} T \text{  for } p \text{ near } p_\text{opt} \text{ (Strange metal)}.
\end{array}
\right.
$$
\vspace{0.5cm}
A linear-in-temperature resistivity is obtained when the doping concentration remains low enough that the quadratic Fermi-liquid term, $U n^2$, is negligible. However, the strange metal phase does not dominate the entire underdoped region beyond the superconducting dome; rather, it competes with the pseudogap phase. In contrast to the strange metal behavior, which aligns naturally with the theory, the pseudogap phase, characterized by a reduction of the density of states near the Fermi level, cannot arise directly from Hund's pairing. Instead, the Hund's coupling tends to enhance metallicity rather than inducing a pseudo-insulating state. Furthermore, the magnitude of the pseudogap systematically decreases with increasing doping, whereas the Hund's parameter may slightly increase with carrier concentration. Consequently, within this theory, the pseudogap is treated as a competing phase rather than a precursor state that generates pairs prior to condensation. This competing order is more likely driven by antiferromagnetic superexchange interactions \cite{Niestemski2009}, which systematically weaken with doping \cite{anderson2002}. Indeed, the superexchange coupling is intimately linked to the localized antiferromagnetic phase that borders the pseudogap regime. Singlet configurations formed via superexchange are inherently less mobile (they are more mobile via doping \cite{Anderson2004}), a feature that may effectively reduce the electronic density of states near the Fermi level in the underdoped region.

Given that the superexchange depends directly on the hopping integral $t$ and the onsite Coulomb parameter $U$, it can modulate the threshold doping parameter $p_\text{min}$ introduced in the model for the superconducting dome. When an isotope substitution occurs, the change in isotopic mass slightly alters the lattice parameters through zero-point phononic modifications. This structural shift induces an oscillation in the superexchange coupling while leaving the purely atomic Hund's parameter largely unaffected. However this influence on $J_\text{SE}$ is weak \cite{Eremin2004,Hfliger2014}, but this slight variation in superexchange alters the stability of the competing antiferromagnetic phase, thereby shifting $p_\text{min}$. Consequently, at a fixed doping level in the underdoped region, the transition temperature can experience either an enhancement or a reduction driven by this parameter shift, whereas the parabolic profile of the dome renders such shifts negligible near $p_{opt}$.

\section{Conclusion}
In summary, the Hund's coupling stands out as one of the few parameters capable of driving an attractive pairing interaction within an otherwise purely repulsive electronic theory of unconventional superconductivity. This mechanism, here denoted as Hund's pairing,  successfully reproduces the experimentally accessible transition temperatures of both cuprate and iron pnictide families within a weak-coupling limit. The attractive contribution mediated by Hund's exchange increases linearly with the doping concentration, whereas the competing Fermi-liquid repulsion scales quadratically. Consequently, a fractional carrier density is required to establish a net attractive binding energy, distinguishing this mechanism from conventional phonon-mediated systems. Furthermore, while unconventional superconductors are intrinsically multi-band systems, a single-band model was strategically implemented here to simplify the approach and capture the fundamental doping asymmetry observed in cuprates. The resulting analytical expression for the superconducting dome achieves quantitative agreement with established empirical profiles using physically realistic values for the Hund's exchange $J$ and onsite Coulomb repulsion $U$. Additionally, the strange metal phase is  captured by the model, where quadratic repulsive interactions become negligible. Conversely, the pseudogap phase does not directly emerge from this pairing theory; it is instead characterized as a competing order likely induced by antiferromagnetic superexchange interactions. This background superexchange coupling may also account for the isotope effects experimentally documented in the underdoped regime.


\bibliography{biblio}
\end{document}